\documentclass{llncs}
\usepackage{url}
\usepackage{hyperref}
\usepackage{epsfig}
\usepackage{graphicx}

\usepackage{todonotes}

\begin{document}


\title{Big Data is not the New Oil: Common Misconceptions about
       Population Data}

\author{Peter Christen\inst{1} \and Rainer Schnell\inst{2}
\institute{
          School of Computing,
          The Australian National University, \\
          Canberra, ACT 2600, Australia.
          \email{peter.christen@anu.edu.au}
          \and
          Methodology Research Group, 
          University Duisburg-Essen, Germany.
          \email{rainer.schnell@uni-due.de}}}

\maketitle


\begin{abstract}
Databases covering all individuals of a population are increasingly
used for research and decision-making. The massive size of such
databases is often mistaken as a guarantee for valid inferences.
However, population data have characteristics that make them
challenging to use. Various assumptions on population coverage and
data quality are commonly made, including how such data were
captured and what types of processing have been applied to them.
Furthermore, the full potential of population data can often only
be unlocked when such data are linked to other databases. Record
linkage often implies subtle technical problems, which are easily
missed. We discuss a diverse range of misconceptions relevant for 
anybody capturing, processing, linking, or analysing population
data. Remarkably many of these misconceptions are due to the social
nature of data collections and are therefore missed by purely
technical accounts of data processing. Many of these misconceptions
are also not well documented in scientific publications. We conclude
with a set of recommendations for using population data.


\end{abstract}


\section{Introduction}

Many domains of science increasingly use large administrative or
operational databases that cover whole populations to replace -- or
at least enrich -- traditional data collection methods such as
surveys or
experiments~\cite{Bradley2021nature,Einav2014science,Foster2017crc}.
This kind of data are now seen as a crucial strategic resource
for research~\cite{Connelly2016ssr,Jorm2015phrp,Porter2020plosone}.
Governments and businesses have also recognised the value that
large population databases can provide to improve
decision-making~\cite{Athey2017science,Isaak2018computer,Provost2013oreilly}.

Due to the perceived advantages of population data, the number of
projects adopting existing databases for research and planning is
increasing. The use of buzzwords like Big data, AI, and machine 
learning, in the context of population data seems to suggest for
non-technical users and decision makers that any kind of question
can be answered when analysing population
databases~\cite{Hand2018rssa}.

However, often neither data quality issues (how population data
were captured and processed) nor the techniques used to link
population data, are clear to decision makers and researchers used
to smaller data sets. Although there is much work on general data
quality, very little specific to population data is published or
taught in data science courses.

Therefore, the kind of problems we consider in this paper are
usually underestimated by non-specialists, leading to inflated
expectations. Such over-expectations might cause costly
mismanagement in areas such as public health or in government
decision-making. Furthermore, failing population data projects,
such as census operations or health surveillance, might even
result in the loss of trust in governments and science by the
public~\cite{Braithwaite2020ajsi,Isaak2018computer}.
%
In the context of research, misconceptions about population data
can lead to wrong outcomes of research studies
that can result in conclusions with severe negative
impact~\cite{Galaitsi2021ijim,Giest2020ps}.
%




%
We focus on personal data about individuals covering (nearly) whole
populations. Following a recent definition of \emph{Population Data
Science}~\cite{Mcgrail2018ijpds}, 
we define population data as \emph{data about people at the level of
a population}. The focus on populations is important, as it refers
to the scale and complexity of the data being considered. These make
(manual) processing and assessment of data quality that are normally
conducted on the much smaller data sets used in traditional medical
studies or social science surveys challenging.

%

Personal data include personally identifiable information
(PII)~\cite{Christen2020springer}, such as the names, addresses, or
dates of birth of people. 
Most administrative and operational data collected by governments
and businesses can also be categorised as personal data, including
electronic health records, (online) shopping baskets, and people's
educational, social security, financial, and taxation
data~\cite{Harron2017bds}.


A crucial aspect of population data is that they are not primarily
collected for research, but rather for operational or
administrative
purposes~\cite{Hand2018rssa,Jorm2015phrp,Keusch2021hcss,Mcgrail2018ijpds}.
As a result, researchers have much less control over such data
and how they are processed, and 
only limited ways to learn about the data's provenance,
making it unclear if such data are fit for the purpose of a
specific research study~\cite{Biemer2017eai}.
%
%
Quoting Brown et al.~\cite{Brown2018pnas}, ``in science, three
things matter: the data, the methods used to collect the data
(which give them their probative value), and the logic connecting
the data and methods to conclusions''. When both the data and their
collection are outside the control of a researcher, conducting
proper science can become challenging. 

Most of the discussion about the use of population data for
research has been about privacy and
confidentiality~\cite{Acquisti2015science,Christen2020springer,Horvitz2015science}.
Much less consideration has been given to how data quality and
assumptions about such data can influence the outcomes of a
research study. For administrative data, decades of experience
have shown that most unusual patterns in large databases are due
to data errors~\cite{Hand2018rssa}. The same can be said about
population covering databases of personal data.

Not much has been written about the characteristics of personal
data and how they differ from other types of data, such as
scientific or financial data.
During the writing of this article, we conducted extensive
literature research to find scientific publications, government
reports, or technical white papers that describe experiences or
challenges when dealing with population databases. The number of
identified publications was scarce, indicating that many challenges
encountered and lessons learnt are not being shared, even though
these would be of high value to 
anybody who works with population data.

Many of the misconceptions
we discuss below are therefore drawn from our several decades of
experience working with large real-world population databases and
in collaborative projects with both private and public sector
organisations across diverse disciplines (including health, finance,
and national statistics).

We do not advocate to abandon the use of population data for
research or decision making. Rather, with this work we aim to
improve the handling of population databases. By demonstrating
common misconceptions about population data, we hope to show how
researchers could identify and avoid the resulting traps.


\begin{figure*}[t]
\begin{center}
  \includegraphics[width=0.99\textwidth]{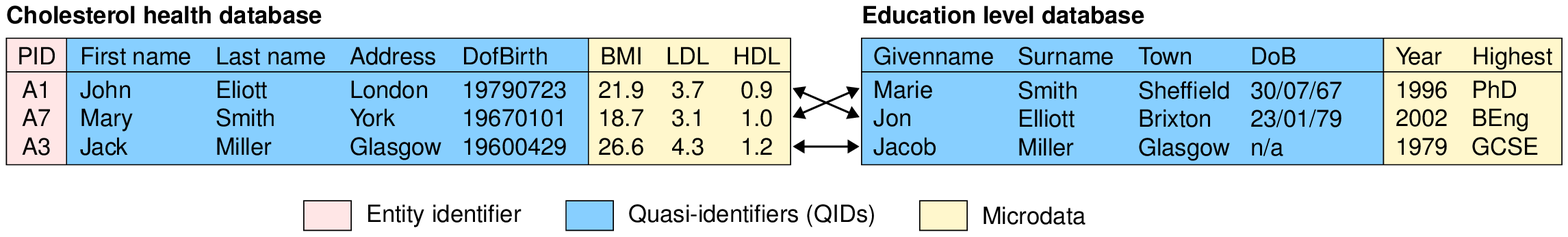}
\end{center}
\caption{
  Two small example population databases, where only one contains
  unique entity identifiers (the PID attribute), but both contain
  quasi-identifiers (QIDs) and microdata. The first database
  consists of eight attributes and the second of six, where both
  have three records. The QIDs have been used to link records
  across these two databases. Note the variations and missing
  values in QIDs.
 \label{fig:data-table}}
\end{figure*}

\section{Characteristics and Uses of Population Data}

Population data are observational data that most commonly occur in
administrative or operational databases as held by government or
business organisations~\cite{Hand2018rssa}.
As we illustrate in Fig.~\ref{fig:data-table}, in their most general
form each \emph{entity} (person) in a population is represented by
one or more \emph{records} (like rows in a spreadsheet) in such a
database,
each consisting of a set of \emph{attributes} (or
\emph{variables}).

The attributes that represent entities in population data can be
categorised into \emph{identifiers} and \emph{microdata}. The first
category can either be an \emph{entity identifier} (such as patient
identifiers) that is unique for each person in a
population~\cite{Christen2012springer}. Alternatively they can be a
group of attributes that, when enough are combined, become unique for
each individual.
Such attributes are known as \emph{quasi-identifiers} (QIDs), and they
include names, addresses, date and place of birth, and so
on~\cite{Christen2020springer}.
Many of the misconceptions about population data are about QIDs and
how they are captured, processed, and used to link 
population databases to transform them into a form suitable for a
research study.

The second component of personal data are known as
\emph{microdata}~\cite{Christen2020springer}, and they
include the data of interest for research, such as the medical,
educational, financial, or location details of individuals. Much
of these data are highly sensitive if they are connected to QID
values because combined they can reveal private personal details
about an individual. Research into data anonymisation and disclosure
control~\cite{Duncan2011book,Elliot2020book} is addressing
how sensitive microdata can be made available to researchers in
anonymised form.

It is commonly recognised that isolated population databases
are of limited use when trying to investigate and solve today's
complex challenges~\cite{Biemer2017eai,Mcgrail2018ijpds}, such as
how a pandemic spreads through a population. Therefore, many projects
that are based on population data employ \emph{record} or \emph{data
linkage}~\cite{Christen2012springer,Harron2015wiley} to link all
records that correspond to the same person across two or more
databases. Linked records at the level of individuals, rather than
aggregated data, are generally required to allow the development of
accurate predictive models.

\begin{figure*}[t]
\begin{center}
  \includegraphics[width=0.95\textwidth]{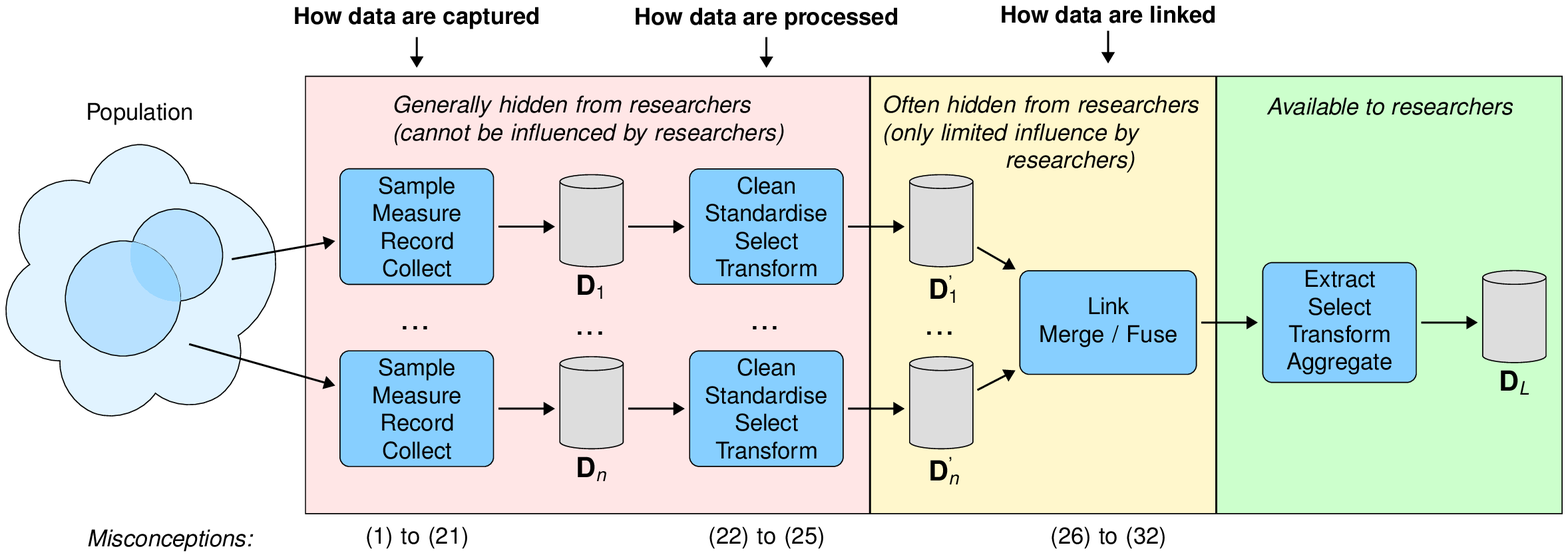}
\end{center}
\caption{
  The general pathway of population data from their source to a
  researcher's computer. We assume $\mathbf{D}_1$ to $\mathbf{D}_n$
  are the source databases that likely cover different subsets of a
  real population. These databases are generally processed
  independently by their respective data owners into the databases
  $\mathbf{D}_{1}^{'}$ to $\mathbf{D}_{n}^{'}$, and then linked and
  processed further (often by different organisations) into the
  linked database $\mathbf{D}_L$ to make it fit for the purpose of
  a research study.\label{fig:data-flow}}
\end{figure*}

We illustrate the general pathway of population data in
Fig.~\ref{fig:data-flow}, and discuss each stage (how
data are captured, processed, and linked) and the corresponding
misconceptions below. While many of these misconceptions
seem obvious, they are often not taken into consideration when
population data are used in research studies or for decision
making.

Many data issues are due to humans being involved in the processes
that generate population data, including the mistakes and choices
people make, changing requirements, novel computing and data entry
systems, limited resources and time, as well as decision making
influenced by political or for economical reasons.
Research managers, and policy and decision makers, often also
assume any kind of question can be answered with highly accurate
and unbiased results when analysing
population databases~\cite{Biemer2017eai}.

While the literature on general data quality is
broad~\cite{Batini2016springer,Kim2003dmkd}, given the
widespread use of personal data at the level of populations it is
surprising that only little published work seems to discuss data
quality aspects specific to personal
data~\cite{Christen2020springer,Smith2018jamia,Tufics2021jdiq}.
One reason is due to the perceived sensitivity of this kind of data:
Population data are generally covered by privacy regulations, such
as the European General Data Protection Regulation 
or the US Health Insurance Portability and Accountability
Act~\cite{Christen2020springer},
and the processes and methods
employed are often 
covered by confidentiality agreements.
%
Furthermore, detailed data aspects are generally not included in
scientific publications where the focus is on presenting the results
obtained in a research study rather than the steps taken to obtain
these results.


\section{Misconceptions about Population Data}


We categorise the identified misconceptions due to how population
data are captured, how such data are processed, and how they are
linked.
We do not discuss any misconceptions related to the analysis of
population data -- how to prevent pitfalls in statistical data
analysis and machine learning has been discussed extensively
elsewhere~\cite{Hastie2009book,Riley2019nature}.

\subsection{Misconceptions due to data capturing}

Misconceptions under this category can occur due to how information
about people is captured. We refer to data capture as any processes
and methods that convert information from a source into electronic
format. This involves decisions about selecting or sampling
individuals from an actual real population to be included into a
population database, how to measure the characteristics of these
people, and the methods employed to actually collect and record this
information.

Many data capturing methods and processes involve humans who can
make mistakes or behave in unexpected ways, or equipment that can
malfunction or be misconfigured. Data capturing methods include
manual data entry, optical character recognition, automatic speech
recognition, and sensor readings (like biometrics from fingerprint
readers and smart watches, or location traces from smart phones).
While each of these 
methods can introduce specific data quality issues (such as keyboard
typing or scanning errors), there are some common misconceptions
about capturing population data.

\smallskip
\textbf{(1) A population database contains all individuals in a
population.}
Even databases that are supposed to cover whole populations, such
as taxation or census databases, very likely have subpopulations
that are under-represented or absent (for example children or
residents without a fixed address). Individuals who do not have a
national health identifier number, like tourists or international
students, and therefore are not eligible for government health
services might be missing from population health databases.
There are also always individuals who refuse to participate in
government services for personal reasons, which can be influenced by
their ethnicity, religion, or political beliefs.

The digital divide~\cite{Dijk2020wiley}, the division of people
who have access to digital services and media versus those who do not,
likely also results in biased population databases. Younger and
more affluent individuals are more likely included in databases
collected using digital services compared to older people,
migrants, or those with a lower socio-economic status.

Organisations might also refuse to provide their data for commercial
reasons or due to confidentiality concerns. Not all patients will be
included in a health database if, for example, certain private
hospitals refuse to participate in a research study. This will likely
introduce bias, since patients with higher income will more likely go
to exclusive private hospitals.

The assumption that all individuals in a population are represented
in a database leads to the illusion that it will be possible to
identify any subpopulation of interest and explore this group of
people, even if it is very small~\cite{Hand2018rssa}.

\smallskip
\textbf{(2) The population covered in a database is well defined.}
The reasons for records about individuals to be included (or not)
into a database are crucial to understanding the population covered
in that database. Some population databases can be based on mandatory
inclusion of individuals (think of government taxation or residency
databases) while others are based on voluntary or self-selected
inclusion (think of health databases such as registries where
patients can opt-out when asked to provide their medical details).

%
The definitions and rules used to extract records about 
individuals into a population database might not be known to those
who are processing and linking the database, and even less likely
to the researchers who will be analysing it~\cite{Biemer2017eai}.
Furthermore, these rules 
might differ between organisations or change over time, or they
might contain minor inconsistencies or mistakes such as ill-defined
age or date ranges.
For example, COVID-19 cases can be included into a database
based on the date of symptom onset, collection of samples, or
diagnosis~\cite{Badker2021bmjgh}, where each of these will
result in different numbers of records being added to a database
each day.



\smallskip
\textbf{(3) Population databases contain complete information for
  all records in the database.}
%
%
%
Many population databases contain different pieces of information
for different sets of records. This sparseness is because large
databases are commonly generated by compiling different individual
databases each only covering a part of a population, or by
collecting records over time where changes in regulations and data
capturing methods and processes can lead to different attributes
being collected. The resulting sparse patterns may prevent many
statistical analyses of the data if not all relevant information has
been captured.


\smallskip
\textbf{(4) All records in a population database are within the
  scope of interest.} 
Individuals might have left the population of interest for a research
study because the criteria for inclusion in a database are no longer
given. For example, a person may have died (although this might be
of interest in itself) or have left the geographical area of a
study. Because many organisations are not notified of an individual's
death or relocation until some time after the event (in some cases
never), at any given time a population database will likely contain
records that are outside the scope of interest.
Including records about these people can affect research studies as
well as operational systems.
%

%

\smallskip
\textbf{(5) Each individual in a population is represented by a
  single record in a database.}
%
A database may contain duplicate records referring to the same
person due to errors or variations in QID values. While data entry
validation may prevent exact duplicates, fuzzy or approximate
duplicates~\cite{Christen2012springer} might be missed by
(automatic) checks. In real-world settings, the same person can
therefore be registered multiple times at different institutions
and their duplicate records are not being identified as referring to
the same individual.

Some duplicates are very difficult to find, for example women who
change their last name and address when they get married and
therefore only their first name, gender, and place and date of birth
stay the same. The flipside is that several people with highly similar
personal details (similar QID values), such as twins who only have
different first names, might not be recognised as two individuals
but instead as duplicates.

Duplicate records are possible even if entity identifiers (such as
social security numbers or patient identifiers) are available that
should prevent multiple records by the same individual from being
added to a population database. Due to human behaviour and errors
such identifiers are not always provided or entered correctly.

An example are people in a social security database who should have
one record but were registered multiple times because they changed
their names or addresses and might have forgotten their previous
registration (and their unique identifier number), or they might be
interested in multiple registrations to obtain social benefits more
than once.
%
%
Duplicate records have even been identified in domains
where high data quality is crucial, such as in voter registration
databases~\cite{Panse2021edbt}.


\smallskip
\textbf{(6) Records in a population database always refer to real
people.}
Real-world databases can contain records of people who never
existed~\cite{Christen2016jdiq}. These can be records added
to train data entry personnel or test software systems. Often these
records are not removed from a database. While potentially easy to
identify by humans (`Tony Test' living in `Testville'), these
records are difficult to detect by data cleaning algorithms because
they were designed to have the characteristics of real people and
often contain high amounts of variations and errors.

In population databases collected from social media platforms,
complete records might furthermore correspond to fake users, and
increasingly even AI (artificial intelligence) bots that generate
human-like content~\cite{Biemer2017eai,Keusch2021hcss}.

\smallskip
\textbf{(7) Errors in personal data are not intentional.}
There are social, cultural, as well as personal reasons why
individuals would decide to provide incorrect personal details.
These include fear of surveillance by governments, trying to prevent
unsolicited advertisements from businesses, or simply the desire to
keep sensitive personal data private.
Fear of data breaches and how personal data are being (mis)used by
organisations are clearly influencing the reluctance of individuals
to provide their details unless deemed
necessary~\cite{Isaak2018computer,Siegel2013wiley}. In
domains such as policing and criminal justice, faked QID values
such as name aliases occur commonly as criminals try to hide their
actual identities~\cite{Larney2011er}.

Incorrectly provided data might only be modified slightly from a
correct value (such as a date of birth a few days off), be changed
completely (a different occupation given), not be provided at all
(no value entered if an input field is not mandatory), or be made
up (such as a telephone number in the form of `1234 5678').
%

The decision to provide incorrect or withhold personal details is
dependent upon where this information is being collected. It is less
likely for an individual to provide incorrect values on an official
government form or when opening a bank account compared to when
ordering a book in an online store.

\smallskip
\textbf{(8) Certain personal details do not change over time.}
While some personal details, such as names and addresses, are known
to change over time for many individuals, it is often assumed that
others are fixed at birth. These include ethnic and gender
identification, as well as place and country of birth.
In many population databases, ethnic identification is
self-reported, where the available categories depend upon how a
society values different subpopulations. For example, the US Bureau
of the Census has repeatedly changed the details of their policy
regarding ethnic
groups\footnote{\url{https://www.census.gov/about/our-research/race-ethnicity.html}}.
Socially influenced attributes might be changed by individuals over
time, a recent example being the Black Lives Matter movement which
made many individuals become more proud to be of colour. It can
therefore be problematic to use these values in the context of, for
example, longitudinal data analysis or record
linkage~\cite{Christen2020springer}.

It is even possible for values fixed at birth to change in the
real world. An example is the Eastern German city of Chemnitz,
which from 1953 until 1990 was named Karl-Marx-Stadt. Individuals
born during that period have a country of birth (German Democratic
Republic) and a place of birth that both do not exist anymore.

\smallskip
\textbf{(9) Personal name variations are incorrect.}
%
%
People's names are a key component of QIDs in many population
databases.
Unlike with most general words, for many personal names there are
multiple spelling variations (such as `Gail', `Gayle', and `Gale'),
and all of them are correct~\cite{Christen2012springer}. When data
are entered, for example over the telephone, differently sounding
name variations might be recorded for the same individual due to
mispronunciation or misunderstanding.

Furthermore, there are many cultural aspects of names, including
different name structures, ambiguous transliterations from non-roman
into the roman alphabet, or name changes over time for religious
reasons, to name a few. Name variations are a known problem when
names are compared between records when linking
databases~\cite{Herzog2007springer}. Working with names can
therefore be a challenging undertaking that requires expertise in
the cultural and ethnic aspects of
names~\cite{Mckenzie2010falsehoods}.

\smallskip
\textbf{(10) Coding systems do not change over time.}
Categorical QID values and microdata are often coded using systems
such as the International Standard Classification
of Occupations
(ISCO)\footnote{\url{https://www.ilo.org/public/english/bureau/stat/isco/}}
or the International Classification of Diseases
(ICD)\footnote{\url{https://www.who.int/standards/classifications/classification-of-diseases}},
the latter currently in its eleventh revision. It is commonly
assumed that such codes are fixed over time and unique in that a
certain item, such as an occupation or disease, is only assigned
one code, and that this assignment does not change.
However, many coding systems are revised over time with new codes
being added, outdated and unused codes being removed, and whole
groups of codes being re-coded (including codes being swapped). A
database might therefore contain codes which are no longer valid.

An example are the codes of the Australian Pharmaceutical Benefit
Scheme (PBS)~\cite{Mellish2015biomed}, where the antidepressant
Venlafaxine had the code N06AE06 until 1995, when it was given the
code N06AA22, which was then changed to N06AX16 in 1999.
%
Using such codes to group or categorise records can therefore lead
to wrong results of a research study  if records have been collected
over time.



\smallskip
\textbf{(11) Data definitions are unambiguous.}
Many population databases contain information that is based on
definitions such as how to categorise records or create categorical
QID or microdata values. As with coding systems, data definitions
can change over time, and they can also be interpreted differently.
A recent example are the definitions for death or hospitalisations
due to COVID-19 infections, where different US states used various
definitions that resulted in databases that could not be used for
comparative analysis~\cite{Galaitsi2021ijim}.


Unless metadata (see misconception 25) are available that clearly
describe such definitions and their changes, it can be difficult to
identify the effects of any changed definitions because any such
change might have subtle effects on the characteristics of only
some individuals in the population of interest for a research study.

\smallskip
\textbf{(12) Temporal data aspects do not matter.}
Given the dynamic nature of personal details, the time and date
when population data are captured and stored in a database can be
crucial because differences in data lag can lead to inconsistent
data that are not suitable for research
studies~\cite{Bradley2021nature,Galaitsi2021ijim}. If it
takes different amounts of time for different organisations to
capture data about the same events then clearly these data are not
comparable, resulting in misreporting for example of the numbers of
daily COVID-19 deaths~\cite{Badker2021bmjgh} or vaccination
rates~\cite{DW2021,Bradley2021nature}. Daily, weekly,
monthly, or seasonal aspects can influence data measurements, as
can events such as public holidays and religious festivities which
likely only affect certain subpopulations.

Data corrections are not uncommon, especially in applications where
there is an urgent need to provide initial data as quickly as
possible, for example to better understand a global
pandemic~\cite{Badker2021bmjgh,Bradley2021nature}. Later
updates and corrections of data might not be considered, leading to
wrong conclusions of research studies.

\smallskip
\textbf{(13) The meaning of data is always known.}
It is not uncommon for population databases to contain attributes
that are not (well) documented. These can include codes without
known meaning, irrelevant sequence numbers, or temporary values
that have been added at some point in time for some specific
purpose. If no documentation is available, database managers are
generally reluctant to remove such attributes. As a result, spurious
patterns might be detected if such attributes are included into a
data analysis.

\smallskip
\textbf{(14) Missing data have no meaning.}
Missing data are common in many databases~\cite{Hand2020pup}.
They can lead to problems with data processing, linking, and
analysis~\cite{Christen2020springer}. Missing data can occur
at the level of missing records (no information is available about
certain individuals in a population), missing attributes (some QIDs
or microdata contain no data values for all records in a database),
or missing QID or microdata values for individual records (specific
missing attribute values)~\cite{Biemer2017eai}.

There are different categories of missing
data~\cite{Hand2020pup,Little2020wiley}. In some cases a
missing value does not contain any valuable information, in others
it can be the only correct value (children under a certain age should
not have an occupation), or it can have multiple interpretations. A
missing value for a question about religion in a census, for example,
can mean an individual does not have a religion or they choose not
to disclose it. Missing data can also occur in settings where
resources are limited and therefore data entries had to be
prioritised, such as in emergency departments~\cite{Badker2021bmjgh}.

Care must therefore be taken when considering missing data. Removing
attributes or even records with missing values, or imputing missing
values~\cite{Herzog2007springer,Little2020wiley}, can result in
errors and structural bias being introduced into a population database
that can lead to incorrect outcomes of a research study.

\smallskip
\textbf{(15) All records in a population database were captured
  using the same process.}
Since population databases are often collected over long periods of
time and wide geographical areas, records are commonly generated or
entered by a large number of staff, giving rise to different
interpretations of data entry rules. For example, if an input field
requires a mandatory value, humans will enter all kinds of
unstandardised indicators for missingness, ranging from single
symbols (like `--' or `.'), acronyms (`NA' or `MD'), to texts
explaining the missing data (like `unknown'). If a database is
compiled from independent organisations
these different interpretations of data entry rules will cause the
need for standardisation before analysis. Manual data entry such as
typing can furthermore lead to different error characteristics
between data entry personnel~\cite{Christen2020springer},
resulting in subsets of records in a population database with
different data quality.

%
Data might be captured at different temporal and spatial resolution,
such as postcodes or city names only versus detailed street
addresses.
%
As a result, the characteristics of both QID and microdata values can
differ between subsets of records in a population database, making
their comparison and analysis challenging~\cite{Badker2021bmjgh}.


\smallskip
\textbf{(16) Attribute values are correct and valid.}
Any data values captured, either by some form of sensor or manually
entered into a database, can be subject to errors coming from
equipment malfunction, human data entry (typing mistake), or
cognitive mistakes (such as confusion about the data required or
difficulties recalling correct information), or even malicious
intent~\cite{Biemer2017eai,Kim2003dmkd}.

In the medical domain, manual typing errors, wrong interpretations
of forms (think of handwritten prescriptions by doctors), entering
values into the wrong input fields, or making mistakes interpreting
instructions (when prescribing drugs) are commonly occurring
mistakes, where rates ranging from 2 to 514 mistakes per 1,000
prescriptions have been reported~\cite{Velo2009bjcp}.

While data validation tools can detect values outside the range or
domain of what is valid (such as 31st
February)~\cite{Kim2003dmkd}, without 
external validation it is generally not feasible to ascertain the
correctness of any given value. If a patient is really 42 years
old can only be validated if
authoritative information (likely from an external database) about
the patient's true age is available.

Furthermore, while individual QID values in a given record can each
be valid, they might contradict each other. For example, a
record with first name `John' and gender 'f' likely contains one
QID value that is incorrect. Many, but not all, such contradictions
can be identified and corrected using appropriate edit
constraints~\cite{Herzog2007springer}.

\smallskip
\textbf{(17) Data values are in their correct attributes.} 
Data entry personnel do not always enter values into the correct
attribute. Many Asian and some Western names, for example, can be
used interchangeably as first and last names, leading to
misinterpretation. For example, `Paul', `Thomas', `Chris', and
`Dennis' are all used as first and last names.

\smallskip
\textbf{(18) Data validation rules produce correct data.}
To ensure data of high quality, many data management systems
contain rules that need to be fulfilled when data are being
captured. For example, registering a new patient in a hospital
requires both a valid address and a valid date of birth. In some
cases, such as in emergency admissions, not all of this information
will be known. Due to such data validation rules, default values are
often used, a common example being the 1 January for individuals
with unknown day and month of birth. While these are valid, if not
handled properly such defaults can result in skewed data distributions
that can adversely affect research studies. Data entry personnel
might also have ad-hoc rules they apply in order to bypass data
entry requirements and to ensure any entered records fulfil all
data validation steps.

\smallskip
\textbf{(19) All relevant data have been captured.}
Because the primary purpose of most population databases is not
their use for research studies,
not all relevant information that is of
importance for a given study might be available for all records in a
database, or it might only be available in subsets of records. This
can, for example, be due to changes in data entry requirements over
time, data might have been withheld by the owner due to
confidentiality concerns or for commercial reasons, or only be
provided in aggregated or anonymised form. If a statistical model
is generated from the data, a probable causal variable might be
missed, since it is not captured at all. 

Data that are not available are known as \emph{dark
data}~\cite{Hand2020pup}, data we do not know about but that
could be of interest to a research study.
As a result certain required or desired information might be
missing for a given research study, making a given population
database less useful or requiring the use of alternative data for
that study.

\smallskip
\textbf{(20) Population data provide the same answers as survey
data.}
Population data, as captured from administrative or operational
databases, are about what people are and what they
do~\cite{Hand2018rssa}. This is unlike survey data where commonly
questions about attitudes, beliefs, expectations, or intentions are
asked with the aim to understand the behaviour of people. Factual
information about people can provide different answers compared to
questions about what they claim to do, while inferring people's
beliefs from their behaviour might not be possible.

\smallskip
\textbf{(21) Population data are always of value.}
Both private and public sector organisations increasingly make
databases publicly available to facilitate their analysis by
researchers. However, many of these databases either lack metadata
or context for them to be of use, or they are aggregated or
anonymised due to privacy and confidentiality
concerns~\cite{Elliot2020book}. A main reason for this is
because past experiences have shown that sensitive personal
information about individuals can sometimes be reidentified even
from supposedly anonymised databases~\cite{Siegel2013wiley}.

Population data without context are unlikely to be of use for
research studies. A database of QID values (such as names and
addresses) without any (or only limited) microdata is, by itself, of
little value for research. Having the educational level of
individuals in a database only becomes useful if this database can
be linked with other data at the level of individuals. Furthermore,
due to the dynamic nature of people's lives, population data become
out of date quickly and therefore need to be updated regularly.
Without adequate metadata (see misconception 25), context, useful
detailed microdata, and regular updates, many publicly available
databases are of little value for research.


\subsection{Misconceptions due to data processing}

It is rare for population databases to be used for research without
any processing being conducted. The organisation(s) that collect
population data, and those that further aggregate, link (or otherwise
integrate) such data, as well as the researchers who will analyse the
data, all will likely apply some form of data
processing~\cite{Badker2021bmjgh}.

Processing can include data cleaning and standardisation, parsing
of free text values, transformation of values, 
numerical normalisation, re-coding into categories, imputation of
missing values, and data aggregation.
The use of different database management systems and data analysis
software can furthermore result in data being reformatted internally
before being stored and later extracted for further processing and
analysis.
Each component of a data pipeline can result in both explicit (user
applied) as well as implicit (internally to software) data processing
being conducted, leading to various misconceptions.

\smallskip
\textbf{(22) Data processing can be fully automated.}
Much of the processing of population data has to be conducted in an
iterative fashion, where data exploration and profiling lead to a
better understanding of a database which in turns helps to
apply appropriate data processing techniques~\cite{Kim2003dmkd}.
This process requires
manual exploration, programming of data specific functionalities,
domain expertise with regard to the provenance and content of a
database, as well understanding of the final use of a database.
Data processing is often the most time-consuming and resource
intensive step of the overall data analytics pipeline, commonly
requiring substantial domain as well as data expertise. 
In national
statistical agencies, it has been reported that as much as 40\% of
resources are used on data processing~\cite{Biemer2017eai}.

Time and resource constraints might mean not all desired data
processing can be accomplished. Manual editing and evaluation is also
unlikely to be possible on large and complex population databases,
and therefore compromises have to be made between data quality and 
timeliness for a database to be made available for
research~\cite{Biemer2017eai}.

\smallskip
\textbf{(23) Data processing is always correct.}
There are often multiple methods available to process data, for
example to normalise numerical values, impute missing data, or
standardise free-format
text~\cite{Batini2016springer,Herzog2007springer}.
Converting `dirty' data into `clean' data can therefore result in
incorrectly cleaned data~\cite{Christen2012springer}. Sometimes
there is no single correct value for a given ambiguous input value.
For example, within a street address, the abbreviation `St' can
stand for either `Street' or `Saint' (as used in a town name like
`Saint Marys').
%

Given data processing commonly involves human efforts, mistakes in
the use and configuration of software can lead to incorrect data
processing, as can bugs in or the use of different or outdated
versions of software. The use of unsuitable tools for a given
project (such as spreadsheet software instead of a proper database
management system or statistical analysis software) can furthermore
results in mistakes when data are being processed. It has been
reported~\cite{Fetzer2021pnas} that on 2 October 2020 a total
of 15,841 positive COVID-19 cases (around 20\%) in England were missed
because when recording daily cases an old file format of the
Microsoft Excel spreadsheet software was used which allowed a
maximum of 65,536 rows. Software features such as auto-completion
and automatic spelling correction can furthermore lead to the wrong
correction of unusual but valid data values that are not available
in a dictionary.

\smallskip
\textbf{(24) Aggregated data are sufficient for research.}
Highly aggregated data, for example at the level of states, counties,
or large geographical units, are hardly of use for scientific
research intending causal statements. Although results based on
aggregated data might seem to be interesting, the number of possible
alternative explanations for the same set of facts based on
aggregated data is usually so large that no definitive conclusions
are possible.

A major problem here is the \emph{ecological fallacy}, describing the
mistake of an aggregate relationship implying the same relationship
for individuals~\cite{Firebaugh2001iesbs}. For example, if
increased mortality rates are observed in regions where vaccination
rates are high, the false conclusion would be that vaccinated people
have a higher probability to die. But actually the reverse might be
true: People observing other people dying might be more willing to
get vaccinated.

How data are aggregated depends on how aggregation functions are
defined and interpreted. 
Weekly counts can, for example, be summed Monday to Sunday or
alternatively Sunday to Saturday. Data that are aggregated
inconsistently, or at different levels of aggregation, will unlikely
be of use for any research studies (or only after additional data
processing has been conducted).

\smallskip
\textbf{(25) Metadata are correct, complete, and up-to-date.}
Metadata (also known as \emph{data dictionaries}) are describing a
database, how it has been created, populated, and its content
captured and processed. Metadata include aspects such as the source,
ownership, and provenance of a database, licensing and access
limitations, costs, description of all attributes including their
domains and any coding systems used, summary statistics and
descriptions of data quality dimensions~\cite{Christen2020springer},
as well as any data cleaning, imputation, editing, processing,
transformation, aggregation, and linkage that was conducted on the
source database(s) to obtain a given population database.
Relevant documentation should be provided, including who conducted
any data processing using what software (and which version of it),
and containing a revision history of that documentation. Metadata are
crucial to understand the actual structure, content, and quality of
a database at hand.

Unfortunately, metadata are often not available, or they are
incomplete, out of date, they need to be purchased, or can only
be obtained through time-consuming approval
processes~\cite{Jorm2015phrp}. A lack of metadata can lead to
misunderstandings during data processing, linking, and analysis,
wasted time, misreporting of results, or can make a population
database altogether useless~\cite{Galaitsi2021ijim}.


\subsection{Misconceptions due to data linkage}

Linking databases is generally based upon comparing the QID values
of individuals, such as people's names, addresses, and other personal
details (as illustrated in Fig.~\ref{fig:data-table}), to find
records that refer to the same
person~\cite{Harron2015wiley,Herzog2007springer}.
These QID values, however, can contain errors, be missing, and they
can change over time. This can lead to incorrect linkage results
even when modern linkage methods are
employed~\cite{Binette2022scadv}. Linking databases can therefore
be the source of various misconceptions about a linked data set.

\smallskip
\textbf{(26) A linked data set corresponds to an actual population.}
Due to data quality issues and the record linkage technique(s)
employed~\cite{Christen2012springer}, a linked data set likely
contains wrong links (Type I errors, two records referring to two
different individuals were linked wrongly) while some true links
have been missed (Type II errors, two records referring to the same
person were not linked)~\cite{Doidge2019ije}. The performance of
most record linkage techniques can be controlled through
parameters~\cite{Binette2022scadv}, allowing a trade-off between
these two types of errors. Many linked data sets with different
error characteristics can therefore be generated by changing
parameter settings, where each provides an approximation of the
actual population it is supposed to represent.

\smallskip
\textbf{(27) Population databases represent the conditions of people
at the same time.}
Data updates on individuals often occur at different points in time,
usually when an event such as a medical condition occurs, or a data
error is detected during a triggered data transaction such as a
payment. In the German Social Security database, for example,
education is entered when a record for a given person is newly created
and not regularly updated afterwards. Therefore, highly trained
professionals might have a record stating a low educational level
because they were pupils at the time of their first paid job.
As a result, records that
represent the same individual can have different values in both QID
and microdata values. The assumption that the QID values of all
records in the databases being linked are up-to-date might therefore
not be correct, and outdated information can lead to wrong linkage
results~\cite{Christen2020springer}.

Data corrections and updates can furthermore occur when incorrect
historical data are being discovered and errors
rectified~\cite{Bradley2021nature}. Unless it is possible to
re-conduct a linkage, which is unlikely for many research studies
due to the efforts and costs involved in such a process, a linked
data set might contain errors which have influenced the conclusions
of the original study.

\smallskip
\textbf{(28) A linked data set contains no duplicates.}
When linking databases, pairs or groups of records that refer to the
same individual might not be linked correctly (missed true links,
Type II errors). One reason for this to occur is if a wrong entity
identifier has been assigned to an individual, as has been reported
even in voter databases~\cite{Panse2021edbt}. Another reason is
if 
crucial QID values of an individual have changed over time, such as
both their name and address details, or are missing, resulting in
two records that are not similar with each
other~\cite{Christen2012springer}.
Therefore, many linked data sets do contain more than one record for
some individuals in a population.

\smallskip
\textbf{(29) A linked data set is unbiased.}
Linkage errors generally do not occur at
random~\cite{Jorm2015phrp}, rather they depend upon the
characteristics of the actual QID values of individuals, which can
differ in diverse subpopulations. Examples include name structures
of migrants that are different from the traditional Western standard
of first, middle, and last name formats~\cite{Mcgrath2021ldh},
or different rates of mobility (address changes) for young versus
older people. As a result, there can be structural bias in a linked
data set in subpopulations defined by ethnic or social categories,
age, or gender (for example if women are more likely to change their
names compared to men when they get
married)~\cite{Bohensky2010bmchsr}.

Recent work has also shown that even small amounts of linkage error
can result in large effects on false negative (Type II) error rates
in research studies. This is especially the case with small sample
sizes that can occur with the rare effects that are often sought to
be identified via record linkage from large population
databases~\cite{Tahamont2021jqc}.
%
%
If the aim of a study is to analyse certain (potentially small)
subpopulations, or
compare, for example, health aspects between subpopulations, then a
careful assessment of the potential bias introduced via record
linkage is of crucial importance~\cite{Doidge2019ije}.

\smallskip
\textbf{(30) Attribute values in linked records are correct.}
%
Given a supposedly correct link, there might be contradicting
attribute values in the corresponding records. Data fusion is the
process of resolving such inconsistencies, where often a decision
needs to be made which of many available fusion operations to
apply~\cite{Bleiholder2008acmcs}. Even if the links made
between records are correct, how records are fused or merged can
therefore introduce errors both into QID values as well as microdata.

For example, assume three records that refer to the same person have
been linked correctly, where each record contains a different salary
value. Should the average, median, minimum, maximum, or the most
recent of these three salary values be used for the fused record of
this individual? How data fusion is conducted needs to be discussed
with the researchers who will be analysing a linked data set because
depending upon the fusion operation applied substantially different
outcomes will potentially be obtained.

\smallskip
\textbf{(31) Linkage error rates are independent of database size.}
The QID values used to link records can be shared by multiple
individuals, potentially thousands in the case of city and town
names or with popular first and last names. Therefore, when larger
databases are being linked, the number of record pairs with the
same QID values likely increases, resulting in more highly similar
record pairs. Making correct classification becomes increasingly
challenging as there are more possibly matching pairs. Generalising
linkage quality results obtained on small data sets in published
studies to much larger population sized real-world databases can
therefore be dangerous.

\smallskip
\textbf{(32) Modern record linkage techniques can handle databases
  of any sizes.}
Many researchers, especially in the computer science and statistical
domains, who develop record linkage techniques do not have access to
large real-world databases due to the sensitive nature of
population data. As a result, novel linkage techniques are often
evaluated on small public benchmark data sets or on synthetically
generated data~\cite{Christen2020springer}. While error rates for
linkages obtained on such data sets can provide evidence of the
superiority of a novel technique over existing methods, assuming that
this new technique will produce comparable high quality linkage
results on larger real-world databases is not guaranteed. 



\section{Conclusions and Recommendations}

Due to misconceptions such as the ones we have discussed, the much
hyped promise of Big data
requires some careful considerations when personal data at the
level of populations are used for research studies or decision
making. 
%
Given population data
are increasing used in many domains of science,
as illustrated in Fig.~\ref{fig:data-flow}, researchers will
potentially have less and less control over the quality of the data
they are using for their studies and any processing done on these
data~\cite{Brown2018pnas}. They likely will also have only limited
information about the provenance and other metadata that is needed
to fully understand the characteristics and quality of their data.
Because population data are commonly sourced from organisations
other than where they are being
analysed~\cite{Biemer2017eai,Hand2018rssa}, these limitations are
inherent to this kind of data.

There are no (simple) technical solutions to detect and correct many
of the misconceptions we have discussed. What is required is
heightened awareness by anybody working with population data. While
our list of misconceptions is unlikely to be exhaustive,
our aim was to show that there is a broad range of issues that can
lead to misconceptions.
The following recommendations might help to recognise and overcome
such potential misconceptions.

\begin{itemize}
\item If possible, data scientists and researchers should aim to
  get involved in the capturing, processing, and linking of any
  data they plan to use for their research. This involves
  discussions with database owners about what data to collect in
  what format, how to ensure high quality of these data, and that
  adequate metadata are collected.


\item Data scientists and IT personnel who are processing and
  linking population data need to work in close collaboration
  with the researchers who will conduct the actual analysis of
  these data. Forming multi-disciplinary teams with members
  skilled in data science, statistics, domain expertise, as well
  as `business' aspects of research~\cite{Jorm2015phrp}, is
  crucial for successful projects that rely upon population data.
%
  Interaction between data and domain experts might mean that a
  project based on population data becomes an iterative endeavour
  where data might have to be re-captured, re-processed, and
  re-linked until they are suitable for a research study.

%

\item Cross-disciplinary training should be aimed at
  improving complementary skills~\cite{Jorm2015phrp}. Having data
  scientists
  who also have domain specific expertise will be highly valuable in
  any project involving population data.
  Equally crucial is for any researcher, no matter what their
  domain, to understand how modern data processing, record linkage,
  and data analytics methods work,
  and how these methods might introduce bias and errors into the
  data they are using for their research studies. Training in data
  exploration and cleaning 
  methods as well as data quality issues should be part of any degree
  that deals with data, including statistics, quantitative social
  science, computer science, and public health.

\item While extensive methodologies about how to deal with
  uncertainties, bias, and data quality in surveys have been
  developed~\cite{Blair2014book,Reid2017jos}, there is a lack of
  corresponding rigorous methods that can be employed on large
  population databases.
  The Big data paradox~\cite{Meng2018aas}, the illusion that large
  databases automatically mean valid results, requires new
  statistical techniques to be developed.
  %
  While certain data quality issues can be identified
  (and potentially corrected) automatically~\cite{Kim2003dmkd},
  novel data exploration methods are needed to identify more subtle
  data issues where traditional methods are inadequate.

\item A crucial aspect is to have detailed metadata about
  a population database available, including how the database was
  captured, and any processing and linkage applied to it. All
  relevant data definitions need to be described, and information
  about all sources and types of uncertainties need to be
  collected~\cite{Hand2018rssa}. Detailed data profiling and
  exploration should be conducted by researchers before a
  population database is being analysed so any unexpected
  characteristics in their data can be identified.

\item Existing guidelines and checklists, such as
  RECORD~\cite{Benchimol2015plos}
  and GUILD~\cite{Gilbert2017jph},
  should be employed 
  and adapted to other research domains. Frameworks such as the
  Big Data Total Error 
  method~\cite{Biemer2017eai} can be adapted for population
  data to better characterise errors in such data. Furthermore,
  data management principles such as FAIR (Findable, Accessible,
  Interoperable, Reusable)~\cite{Wilkinson2016science} should be
  adhered to, although in some situations the sensitive nature of
  personal data~\cite{Christen2020springer} might limit or prevent
  such principles from being applied.
  In such situations, at least metadata and any software used in a
  study should be made public in an open research repository.
  %

\item The lack of publications that describe practical challenges
  when dealing with population data can result in the
  misconceptions we have discussed here. We therefore encourage
  increased publication of data issues and the sharing of experiences
  with the scientific community about lessons learnt as well as best
  practice approaches being implemented
  when dealing with population data.
\end{itemize}

We have discussed some aspects in modern scientific processes that
are rarely considered when population data are being used for
research studies or decision making. Since good data management
is a key aspect of good
science~\cite{Brown2017biomed,Wilkinson2016science}, it is
vital for anybody who uses population data to be aware of
underlying assumptions concerning this kind of data.
%
We hope the misconceptions and recommendations given here will help
to identify and prevent misleading conclusions and poor real-world
decisions, making population data the new oil of the Big data era.



\subsubsection*{Acknowledgements}

We like to thank S. Bender, S. Redlich, J. Reinhold, and C.
Nanayakkara for their critical and helpful comments, and A.
Pl\"oger for help with producing the figures. P. Christen likes to
acknowledge the support of the University of Leipzig and ScaDS.AI,
Germany, where parts of this work was conducted while he was funded
by the Leibniz Visiting Professorship; and also the support of the
Scottish Centre for Administrative Data Research (SCADR) in
Edinburgh. R. Schnell work's was supported by the Deutsche
Forschungsgemeinschaft Grant 407023611.

\bibliographystyle{splncs04}

\bibliography{paper}


\end{document}